\documentclass[nofootinbib,aps,a4paper,10pt,letterpaper,superscriptaddress,twocolumn,showkeys,times,eqsecnum]{revtex4-2}
\usepackage{amsmath}
\usepackage{amsfonts}
\usepackage{graphicx}
\usepackage{color}
\usepackage{braket}
\usepackage{dcolumn}
\usepackage{bm,url}
\usepackage[linktocpage]{hyperref}
\usepackage{subfigure}
\usepackage{amsfonts}
\usepackage[left=2.5cm,right=2.5cm,top=2cm,bottom=2cm]{geometry}
\usepackage[usenames,dvipsnames,svgnames]{xcolor}  
\usepackage{hyperref}   
\definecolor{oxfordblue}{rgb}{0.0, 0.13, 0.28}
\definecolor{burgundy}{rgb}{0.5, 0.0, 0.13}
\definecolor{darkolivegreen}{rgb}{0.33, 0.42, 0.18}
\definecolor{darkblue}{rgb}{0,0,0.5}
\definecolor{richcarmine}{rgb}{0.84, 0.0, 0.25}
\definecolor{darkblue}{rgb}{0,0,0.5}
\definecolor{bluer}{rgb}{0.00,0.50,0.75}{}
\hypersetup{colorlinks=true, citecolor=cyan, linkcolor=blue,
urlcolor = magenta, filecolor=magenta}

\begin{document}

\newcommand\be{\begin{equation}}
\newcommand\ee{\end{equation}}
\newcommand\bea{\begin{eqnarray}}
\newcommand\eea{\end{eqnarray}}
\newcommand\bseq{\begin{subequations}} 
\newcommand\eseq{\end{subequations}}
\newcommand\bcas{\begin{cases}}
\newcommand\ecas{\end{cases}}
\newcommand{\p}{\partial}
\newcommand{\f}{\frac}

\title{A note on the stability of the Cauchy horizon in regular black holes \\
	without mass inflation}
\author{\textbf{Mohsen Khodadi}}
\email{khodadi@kntu.ac.ir}
\affiliation{Department of Physics, K. N. Toosi University of Technology,	P. O. Box 15875-4416, Tehran, Iran}

\author{\textbf{Javad T. Firouzjaee}}
\email{firouzjaee@kntu.ac.ir}
\affiliation{Department of Physics, K. N. Toosi University of Technology,	P. O. Box 15875-4416, Tehran, Iran}
\affiliation{PDAT Laboratory, Department of Physics, K. N. Toosi University of Technology, P. O. Box 15875-4416, Tehran, Iran}
\affiliation{School of Physics, Institute for Research in Fundamental Sciences (IPM), P.O. Box 19395-5531, Tehran, Iran}

\date{\today}
\begin{abstract}
	
Unlike generic models of regular black holes (BHs) with nonzero surface gravity on both Cauchy and event horizons, an inner-degenerate counterpart with zero Cauchy horizon surface gravity was recently proposed. For this regular BH solution with spherical symmetry, we examine the stability of the Cauchy horizon from both classic and quantum mechanics viewpoints. We find that the classical perturbations do not lead to mass inflation at the Cauchy horizon, indicating classically stability, while quantum fluctuations cause it to be unstable quantum mechanically. 
 
\end{abstract}


\maketitle

\section{Introduction}

The study of the structure of spacetime inside a real black hole (BH) is of great theoretical importance. 
 The central singularity and the existence of the Cauchy horizon are two key components whose as accurate as possible understanding may shed light on the mystery of this inaccessible region of spacetime.
It is known that the classical General Relativity (GR) as a theory with deterministic nature can predict the future directed evolution of the spacetime.
  Based on typical BH solutions of General Relativity such as Reissner--Nordstrom and Kerr, it is expected that gravitational collapse is incomplete geodesically so that stops in a timelike singularity enclosed by the Cauchy horizon. Due to the emergence of some pathological behaviors, Roger Penrose tried to curb it in GR by proposing the idea of cosmic censorship in the form of two weak and strong statements. The weak statement explicitly tells that we have to not worry about the presence of singularity in the center of the BH since it is covered by an event horizon, causing it to be out of observer reach \cite{Penrose}.
 This means that, starting
  from these initial conditions, any sufficiently distant observer will neither encounter any singularities nor any effects arising from propagating out
  of (curvature) singularities. 
The strong statement asserts all the physical solutions in GR originating from Einstein's equations are globally hyperbolic, namely, they are unique, continuous, and predictable \cite{R:1978} (see also \cite{S:1979}). Namely, this conjecture, in essence, formalizes the idea that GR must be a physically reasonable theory that can describe every classical event in the universe. More precisely, it forbids the permanent emergence of the Cauchy horizon, rather it is unstable to remnant perturbation fields and eventually collapses into a singularity \cite{Simpson:1973ua}. In other words, if the initial data is perturbed, a Cauchy horizon does not appear in the resulting spacetime but ends at a singularity.
Indeed, the loss of predictability is a pathological consequence of the presence of the Cauchy horizon in spacetime which is cured if it is unstable \cite{Book}. More strictly, it is well-known from the seminal papers \cite{Poisson:1989zz,Poisson:1990eh} (see also \cite{Mc,Ch}) that any asymptotic flat black hole subject to perturbation is prone to trigger a sufficiently catastrophic phenomenon so-called \textit{"mass inflation singularity"} on the Cauchy horizon to make it destruct. The mass inflation appear as metric perturbations which is a phase of exponential growth of the gravitational energy in a neighborhood of the Cauchy (inner) horizon.
This phenomenon indeed is a result of competition between power-law decay ($\propto r^{-n}$) of time-dependent remnant fields in the external regions of black holes, and a mechanism with an exponential blue-shift amplification of fields ($\propto e^{\kappa_- \nu}$) inside BHs, so that finally the latter is dominated over the former \cite{Hod:1997fy,Hod:1998ra,Hod:1999rx}. Note that decay in the form of power law just takes place in the asymptotic flat BH \cite{Gundlach:1993tp,Gundlach:1993tn}, while in the case of de-Sitter it is exponentially \cite{Brady:1996za} (some more recent studies, such as \cite{Dias:2018ynt,Dias:2018etb,Khodadi:2022xtl}, present the details more clearly and informative).

Although there is no exact theorem to prove these two mathematically independent conjectures, it merely serves to preserve the utility of GR in explaining gravitational phenomena. Consequently, despite the GR suffering from a central singularity, Penrose's conjectures allow it to still be considered a deterministic theory of gravity. Note that the validity of the strong Penrose's conjecture guarantees determinism, but the existence of the Cauchy horizon is not the only source that threatens the determinism and predictive power of the theory, see \cite{Azhar:2021whm} for more detailed discussions.

After all, singular solutions that arose from Einstein field equations cannot be suitable mathematical counterparts to addressing astrophysical BHs. GR is widely believed to be an effective framework for describing gravity, meaning that it is impossible to understand the physics of extreme situations such as singularity in spacetime without a quantum theory of gravity. In other words, the formation of singularity within GR is certain and it is expected that the quantum gravity can handle this unpleasant theoretical effect by regularizing singularity \cite{Crowther:2021qij} or by
strong negative pressure in the interior of the cloud in the very late stages of the collapse \cite{Joshi-book}.
More precisely, the regular BHs, in essence, are deformed versions of vacuum solutions of the Einstein field equations in which instead of the central singularity there is an asymptotically de Sitter (energy density asymptotes to non-zero) \cite{Bardeen} or Minkowski core (energy density asymptotes to zero) \cite{Simpson:2018tsi}. Violation of at least one of the energy conditions in regular BHs is usually the price to pay to bypass the singularity \cite{Khodadi:2022dyi,Sajadi:2023ybm}.	Nonlinear electromagnetic is one of the well-motivated frameworks for finding the physical origin of regular BHs. Although it is expected that in regular BHs, the curvature invariants are finite everywhere in spacetime, it does not mean that these BHs are also free of coordinate singularities. It is worth noting that the boundedness of curvature invariants is not the only condition of regularity, but the geodesic completeness should also be taken into account \cite{Pedrotti:2024znu}. 
The presence of an even number of horizons in regular BHs is essential \cite{Carballo-Rubio:2019fnb,Bonanno:2020fgp} such that due to the presence of a non-extremal inner horizon will result in mass inflation instability \cite{Poisson:1989zz,Balbinot:1993rf}. A recent study conducted in Ref. \cite{Bonanno:2022rvo}, clearly shows that, as a special feature of regular BHs, the whole inner core instability scenario is not just dependent on the mass inflation, but the Hawking flux contribution to the final state of the geometry must also be included.
Overall, for regular (non-singular) BHs to be a viable resolution of the singularity issue, the stability of the Cauchy horizon under any generic perturbation will be important \cite{Carballo-Rubio:2018pmi,DiFilippo:2022qkl}. Specifically, for the spherically symmetric spacetime, it is shown that the stability of the Cauchy horizon to satisfy the regularity conditions is vital \cite{Carballo-Rubio:2019fnb,Carballo-Rubio:2019nel}.
Namely, in the spherically symmetric case a stable core including an inner horizon is required of any regular. It is worth noting that this statement is technically not yet conclusively generalizable to rotating BHs, although it may be supported by several proposals. With these discussions, one can say that the stability study of the Cauchy horizon, in essence, has a twofold motivation. First, within the framework of GR for verifying the validity of the strong cosmic censorship conjecture of Penrose, which one can find in papers \cite{Choquet-Bruhat:1969ywq,Mellor:1989ac,Brady:1992cz,Mellor:1992,Brady:1992,Chambers:1994ap,Markovic:1994gy,Cai:1995ux,Cai:1995nt,Chambers:1997ef,Poisson:1997my,Brady:1998au,Cai:1998yp,Hollands:2019whz,Destounis:2020yav}. Second, in the framework of regularized BHs rendering a stable core is a necessity. The check of the latter is our concern in this manuscript.

The main challenge to having regular BH with a stable core is controlling the exponential instability of so-called mass inflation at the Cauchy horizon. Given that this instability is regulated by the surface gravity of the Cauchy horizon, it is usually handled via two mechanisms. One demands a multi-horizon solution endowed with a degenerate Cauchy horizon, qualified by a zero surface gravity \cite{Carballo-Rubio:2022kad,Franzin:2022wai}, and another argues that Hawking radiation can handle instability by replacing mass inflation with a softer polynomial growth \cite{Bonanno:2020fgp,Bonanno:2022jjp,Bonanno:2022rvo}. 
A significant finding about the latter is that the endpoint of the evolution of a small perturbation on the geometry for such solutions is non-stationary due to the production of a large back--reaction \cite{Carballo-Rubio:2022twq}.

In this regard, by restricting ourselves to spherically symmetric spacetimes, let us introduce the regular BH solution with multi--horizon structures written in advanced null coordinates \cite{Carballo-Rubio:2022kad} 
\begin{equation}\label{eq:genmet}
	\text{d}s^2=-F(r)\text{d}v^2+2\text{d}v\text{d}r+r^2\text{d}\Omega^2, 
\end{equation}
with 
\begin{widetext}
\begin{eqnarray}
	F(r)=\frac{\left(r-r_-\right)^{d}(r-r_+)}{
		\left(r-r_-\right)^d(r-r_+) + 2M r^d + [a_2-3r_-(r_++r_-)] r^{d-1}}\,,~~~~~d>2
	\label{E:F(r)-3}.
	\end{eqnarray}
\end{widetext}
Here, $M$, $r_-$, and $r_+$ are respectively the mass, inner (Cauchy), and outer (event) horizons which are subject to the following conditions \footnote{In the regular model at hand, to make the laps function $F(r)$ used of this simplifying assumption that $F(r)$ is a rational function of the radial coordinate, i.e., a fraction function whose numerator and denominator are polynomials of the same degree $n$. The central idea of zero inner surface gravity in this model of regular BH means that the scales $r_-$, and $r_+$ are separable so that GR still works well around $r_+$, and  it is indeed a single root of $F(r)$ while the root $r_-$ must be degenerated with the degree of degeneracy $d>2$. So, it is very clear that by taking into account $n=4$, namely the lowest possible degree for the polynomials in $F(r)$, then $d=3$. In this case, although the coefficients of numerator of $F(r)$ are determinable in terms of $r_{\mp}$, the denominator coefficients cannot be determined in advance i.e., $F(r)=\frac{(r-r_-)^3(r-r_+)}{a_4r^4+a_3r^3+a_2r^2+a_1r+a_0}$. After applying some conditions such as regularity and asymptotic on the $F(r)$, just coefficient $a_2$ remains after a bit of computational manipulation (more details of calculations find in \cite{Carballo-Rubio:2022kad}).}
\begin{equation}
r_-\ll r_+ \sim 2M, \quad   r_-\sim \left|r_+-2M\right|,
	\quad 
	a_2 \gtrsim {9\over 4} r_+ r_-. 
\end{equation} In (\ref{E:F(r)-3}), $d$ is the degree of degeneracy and, in essence, denotes the number of degenerate roots at the Cauchy horizon. Due to appear of a pole at $r>0$ for the case of $d=2$, then the minimum number of degenerate roots must be $d=3$.
The inner surface gravity of this BH solution is zero, i.e., $\kappa_-=-\frac{1}{2}(\frac{dF}{dr})_{r=r_-}=0$, since $r_-$ is not a single root of $F(r)=0$ which address the location of the inner horizon, while $r_+$ is the only root indicating the event horizon. In other words, the regular BH metric at hand belongs to the multi-horizon spacetimes with three degenerate roots at the Cauchy horizon. Namely, the key reason that surface gravity on the Cauchy horizon is zero is that we, in essence, deal with an inner-degenerate metric.

In the next section, we will investigate the stability issue of the Cauchy horizon of the aforementioned regular BH, both classical and quantum-mechanically. To do the former, we draw inspiration from the classical stability analysis performed in \cite{Brady:1998au}. 
The key achievement in Ref. \cite{Brady:1998au} is that in addition to the well--known generic radiation tail falling off exponentially, outgoing modes trigger a backscattering off due to the spacetime curvature which results in a sub--leading order divergent influx at the Cauchy horizon. In the following, we shall argue this does not occur for the underlying regular BH solution, meaning that it is stable classically. 
Finally, by looking for the behavior of quantum fluctuations via extracting the expression of the vacuum expectation value (VEV) $<T_{\mu\nu}>$ related to quantum fields, we shall reveal that the Cauchy horizon of the regular BH solution at hand, is indeed unstable quantum mechanically.

\section{classical stability and quantum instability}
By re-writing the Vaidya metric (\ref{eq:genmet}) in terms of $(t,r)$ coordinates, we have a static, spherically symmetric
BH with such a degenerate Cauchy horizon
\begin{equation}\label{eq:sh}
\text{d}s^2=-F(r)\text{d}t^2+F(r)\text{d}r^2+r^2\text{d}
	\Omega^2~. 
\end{equation}
The BH solution is supposed to be asymptotically flat. By introducing the null coordinates $(u=t-r^*, v=t+r^*)$, where
$r^* =\int F^{-1}(r)dr$, the BH spacetime above can be re-express as
\begin{equation}\label{eq:sh1}
\text{d}s^2=-F(r)\text{d}u\text{d}v+r^2\text{d}
\Omega^2~,
\end{equation} where at the Cauchy and event horizons have $v=\infty$ and $u=\infty$, respectively\footnote{To have regular coordinates at these horizons, it is sufficient one set the Kruskal-like coordinates \cite{Poisson:1990eh}.}. Inspired by the perturbation method released in Ref. \cite{Brady:1998au}, via investigating the evolution of a classical perturbation field labeled with $\boldsymbol{\Psi}$, we survey the stability issue of the Cauchy horizon. If the perturbation field does not generate a divergent flux at the Cauchy horizon measured by a timelike observer, it is then stable. Otherwise, the backreaction of the divergent flux, causes the Cauchy horizon to be unstable. 

For the resulting flux arising from the perturbation field $\boldsymbol{\Psi}$, measuring by any observer with the four-velocity $U^{\mu}$, we have
\begin{equation}\label{eq:sh2}
\mathcal{F}=\boldsymbol{\Psi}_{,\mu} U^{\mu}~,
\end{equation} where $\mathcal{F}$ indeed is the square of the amplitude of flux measured by observers passing the horizons. It is well-known in case of non-degenerate Cauchy horizon ($\kappa_-\neq0$) the amplitude measured by observers radially passing the Cauchy horizon, takes the following form \cite{Brady:1998au}
\begin{equation}\label{eq:sh3}
\mathcal{F}_-\sim\exp(\kappa_- v)\boldsymbol{\Psi}_{,\mu}.
\end{equation} 
Here $\boldsymbol{\Psi}_{,\mu}$ is of great importance since it, in essence, comes from two contributions.  One is due to the power-law behavior of the tail of the perturbation field at late times in the form $v^{-n-1} (n\geq2)$ \cite{Price:1972pw,Gundlach:1993tp, Burko:1997tb}.
The other is arising from the evolution of the outgoing modes $\boldsymbol{\Psi}_{,\mu}\sim \exp(-\kappa_+ u)$ at the event horizon, leading to backscattering adding an extra contribution to the influx along the Cauchy horizon. As a result, the total amplitude measured by observer passing the Cauchy horizon, reads as \footnote{To know about how in Eq. (\ref{eq:sh4}), $u$-dependency of outgoing modes at event horizon changed to $v$-dependency, see \cite{Brady:1998au}. }
\begin{equation}\label{eq:sh4}
\mathcal{F}_-\sim\exp(\kappa_- v)\bigg(v^{-n-1}+ constant \times \exp(-\kappa_+ v)\bigg).
\end{equation}
It is clear that due to the presence of term $\exp(\kappa_- v)$, and this point that $\kappa_->\kappa_+$ for any separable inner and outer horizons, the expression above will diverge at the Cauchy horizon, as $v\rightarrow\infty$. But when the Cauchy horizon turns degenerate, one faces a different position so that $\exp(\kappa_- v)$ replace with $v^{1+1/d}$, meaning that the expression (\ref{eq:sh4}) remains finite at the Cauchy horizon, as $v\rightarrow\infty$. As a result, one can be said that the regular multi-horizon metric under our attention in this manuscript against the classical perturbation field is stable. This is in perfect agreement with the discussion proposed in \cite{Carballo-Rubio:2022kad} which due to the absence of mass inflation at the Cauchy horizon, we deal with a regular BH with a stable core.
  
However, the story is not expected to end here since a natural demand is the stability evaluation of the Cauchy horizon from the view of quantum mechanics.  In other words, by considering the quantum effects arising from the quantization of matter field perturbations, is the Cauchy horizon still stable or not? It can be asked in another way: could quantum effects affect the classical picture of mass inflation? In earlier research such as \cite{Anderson:1993ni,Balbinot:1993rf,Balbinot:1994ee}, this question has been addressed. The perturbation method is a common one to check the stability of the Cauchy horizon. More precisely, if the evolution of a perturbation field results in producing a divergent flux at the Cauchy horizon measured by a timelike observer, then the Cauchy horizon is no longer stable due to the backreaction of the divergent flux \cite{Brady:1998au}.

In this regard, to reveal the role of quantum fluctuations on the stability of Cauchy horizon, it is essential one extract the VEV $<T_{\mu\nu}>$ related to quantum fields. Although it is difficult to do in physical spacetimes, it is well known that it gets easier in the two-dimensional (2D) reduction model i.e.,
\begin{equation}\label{ds}
 \text{d}s^2=-F(r)\text{d}u\text{d}v.
\end{equation} 

An important point of care in choosing a vacuum state comes from the theorem proposed in the seminal paper \cite{Fulling:1978ht}, which stresses the finite values of any Hadamard state at the event horizon.
As a result, Specifying an appropriate vacuum to get the VEV $< T_{\mu\nu}>$ of the quantum
fluctuations, is essential. Given that we are dealing with the static spherically symmetric BHs in asymptotically flat spacetime,
hence we set Unruh and Israel--Hawking--Hartle vacuum states \cite{Unruh:1976db,Hartle:1976tp} which are more astrophysically relevant vacuum states.  While the former is used as a static state to describe quantum fields in the spacetimes of BHs formed by gravitational collapse and  in the cosmological background \citep{Javad-Ellis}, the latter is suitable for describing BHs in thermal equilibrium with the surrounding thermal radiation. In other words, in the Unruh vacuum one deals with a BH radiating into empty space, while in the Israel--Hartle--Hawking vacuum with a BH which is in equilibrium with a bath of thermal radiation.

By considering a conformally invariant scalar field propagating on the aforementioned 2D background (\ref{ds}),  then its renormalized VEV of the stress-energy tensor is read as \cite{Birrell}
\begin{equation}\label{eq:}
<T_{\mu\nu}>=\Theta_{\mu\nu}+X_{\mu\nu}-(48\pi)^{-1}Rg_{\mu\nu}~.
\end{equation}
Here, the non-zero components of $\Theta_{\mu\nu}$ for exiting 2D background reads as
\begin{equation}\label{}
 \Theta_{uu}=\frac{2F(r)F''(r)-F'^2(r)}{192\pi}=\Theta_{vv}~,
\end{equation} where symbols $'$,  and $''$ , denote the first and second derivative with respect to $r$, respectively. In (\ref{eq:}) $X_{\mu\nu}$  depends on the setting of vacuum states.

The expressions related to the components VEV of the stress-energy tensor for Unruh and Israel--Hartle--Hawking states are identical and take the following forms \cite{Cai:1995nt, Birrell} 
\begin{align}\label{eq:Tu}
&<T_{uu}>_{U,IHH}=\bigg(\frac{\kappa_+^2}{48\pi}+\frac{F(r)F''(r)}{96\pi}-\frac{F'^2(r)}{192\pi}\bigg),\\ 
&<T_{vv}>_{U,IHH}=\bigg(\frac{F(r)F''(r)}{96\pi}-\frac{F'^2(r)}{192\pi}\bigg),\\ 
&<T_{uv}>_{U,IHH}=\frac{F(r)F''(r)}{96\pi}=<T_{vu}>_{U,IHH}.
\end{align}
Subscripts ''U'', and '' IHH'' denote Unruh and Israel--Hartle--Hawking vacuum states, respectively. Despite that these two vacuum states guarantee that the stability of the outer (event) horizon is not threatened by the quantum effects of the perturbation field (e.g., see Refs. \cite{Davies:1976ei,Candelas:1980zt,Fawcett:1983dk,Balbinot:1999vg} and also book \cite{Birrell} for more discussions), it is essential to check it. To do this, we have to check the behavior of the above-mentioned three expressions in limit $r\rightarrow r_+$ i.e.,
\begin{align}\label{eq:lim}
&\lim_{r\rightarrow r_+}<T_{uu}>_{U,IHH}=\frac{\kappa_+^2}{48\pi}- \nonumber \\ &\frac{(r_+-r_-)^{2 d}}{192 \pi  \Big(r_+^{d-1} (a_2-3 r_- (r_-+r_+))+2 M	r_+^d\Big)^2},\\ 
&\lim_{r\rightarrow r_+}<T_{vv}>_{U,IHH}=\\ \nonumber
&-\frac{(r_+-r_-)^{2 d}}{192 \pi  \Big(r_+^{d-1} (a_2-3 r_- (r_-+r_+))+2 M	r_+^d\Big)^2},\\ 
&\lim_{r\rightarrow r_+}<T_{uv}>_{U,IHH}=0.
\end{align}
The finite expressions above as $r\rightarrow r_+$, openly imply that the semiclassical flux does not diverge at the event horizon as expected. It means that the event horizon remains stable against the semiclassical effects of the perturbation field. However, by applying the regularity conditions above, this time on the Cauchy horizon i.e., in the limit $r\rightarrow r_-$, one finds that $uu$ and $vv$ components diverge.

 
 Despite that 2D calculations may have notable differences from those in standard 4D, it expects that the regular BH model at hand with degenerate Cauchy horizon is unstable in the presence of quantum fluctuations. In general, unlike the classical perturbation case, $\kappa_-=0$ no longer guarantees that the Cauchy horizon is also stable against perturbations caused by quantized matter fields.


In summary, we have taken one kind of spherically symmetric regular inner-degenerate BH solution in asymptotically flat spacetime with zero surface gravity \cite{Carballo-Rubio:2022kad} in which mass inflation is absent. It is well known that the appearance of mass inflation implies that a scalar curvature singularity must be formed at the Cauchy horizon. However, in the absence of mass inflation, whether or not a scalar curvature singularity forms is a significant theoretical question, which should be investigated from two viewpoints: classically and quantum mechanically. 
We have found that the BH solution at hand is an example of a regular core with classical stability, without forming the scalar curvature singularity on the Cauchy horizon. However, by taking the quantum considerations into field perturbations, the structure of the Cauchy horizon is no longer stable, and quantum mechanically unstable. Namely, the Cauchy horizon in this class of regular black holes (without mass inflation) still is unstable quantum mechanically and converts into a scalar curvature singularity. 

It should be mentioned that at the end of this study, we realized that in Ref. \cite{McMaken:2023uue}, the author has provided a detailed analysis of semiclassical divergence of energy density on the Cauchy horizon of the underlying regular metric,  using the effective Hawking temperature and the renormalized stress-energy tensor. The main result presented in this letter i.e., the lack of quantum mechanical stability, is consistent with \cite{McMaken:2023uue}.

\bigskip
\section* {ACKNOWLEDGMENTS} 
M. Kh thanks Stefano Liberati for clarifying some questions and providing insightful discussions in the early stages of this work.


\newpage
\begin{thebibliography}{99}

\bibitem{Penrose}	
R. Penrose, Riv. Nuovo Cim. 1, 252-276 (1969) [Gen. Rel. Grav. 34, 1141 (2002)].

\bibitem{R:1978}
R. Penrose, \textit{''Singularities of Spacetime, Theoretical Principles in Astrophysics and Relativity''} (A78-43851
19-90), Chicago University Press, Chicago (1978).


\bibitem{S:1979}
S.W. Hawking and W. Israel, \textit{''General Relativity, an Einstein Centenary Survey''}, Cambridge University Press,
Cambridge (1979).


	
	
\bibitem{Simpson:1973ua}
M.~Simpson and R.~Penrose,
Int. J. Theor. Phys. \textbf{7}, 183-197 (1973).

\bibitem{Book}
D. Grumiller and M.M. Sheikh-Jabbari, \textit{''Black Hole Physics
From Collapse to Evaporation''}, Springer, 1st ed. (2022).

\bibitem{Poisson:1989zz}
E.~Poisson and W.~Israel,
Phys. Rev. Lett. \textbf{63}, 1663-1666 (1989).

\bibitem{Poisson:1990eh}
E.~Poisson and W.~Israel,
Phys. Rev. D \textbf{41}, 1796-1809 (1990).

\bibitem{Mc}
J. M. McNamara, 
Proceedings of the Royal Society of London. Series A,
Mathematical and Physical Sciences 358, 499 (1978).

\bibitem{Ch}
S. Chandrasekhar and J. B. Hartle, 
Proceedings of the Royal Society of London. Series A, Mathematical and Physical Sciences 384, 301 (1982).

\bibitem{Hod:1997fy}
S.~Hod and T.~Piran,
Phys. Rev. D \textbf{58}, 024018 (1998)
[arXiv:gr-qc/9801001 [gr-qc]].

\bibitem{Hod:1998ra}
S.~Hod and T.~Piran,
Phys. Rev. D \textbf{58}, 044018 (1998)
[arXiv:gr-qc/9801059 [gr-qc]].

\bibitem{Hod:1999rx}
S.~Hod,
Phys. Rev. D \textbf{61}, 024033 (2000)
[arXiv:gr-qc/9902072 [gr-qc]].

\bibitem{Gundlach:1993tp}
C.~Gundlach, R.~H.~Price and J.~Pullin,
Phys. Rev. D \textbf{49}, 883-889 (1994)
[arXiv:gr-qc/9307009 [gr-qc]].

\bibitem{Gundlach:1993tn}
C.~Gundlach, R.~H.~Price and J.~Pullin,
Phys. Rev. D \textbf{49}, 890-899 (1994)
[arXiv:gr-qc/9307010 [gr-qc]].


\bibitem{Brady:1996za}
P.~R.~Brady, C.~M.~Chambers, W.~Krivan and P.~Laguna,
Phys. Rev. D \textbf{55}, 7538-7545 (1997)
[arXiv:gr-qc/9611056 [gr-qc]].

\bibitem{Dias:2018ynt}
O.~J.~C.~Dias, F.~C.~Eperon, H.~S.~Reall and J.~E.~Santos,
Phys. Rev. D \textbf{97}, no.10, 104060 (2018)
[arXiv:1801.09694 [gr-qc]].

\bibitem{Dias:2018etb}
O.~J.~C.~Dias, H.~S.~Reall and J.~E.~Santos,
JHEP \textbf{10}, 001 (2018)
[arXiv:1808.02895 [gr-qc]].


\bibitem{Khodadi:2022xtl}
M.~Khodadi and J.~T.~Firouzjaee,
Phys. Dark Univ. \textbf{37}, 101084 (2022)
[arXiv:2206.12843 [gr-qc]].

\bibitem{Azhar:2021whm}
F.~Azhar and M.~H.~Namjoo,
[arXiv:2101.10887 [physics.hist-ph]].

\bibitem{Crowther:2021qij}
K.~Crowther and S.~De Haro,
[arXiv:2112.08531 [gr-qc]].	

\bibitem{Joshi-book}
Joshi, Pankaj S. Gravitational collapse and spacetime singularities. 2007.



\bibitem{Bardeen}
J. M. Bardeen, “Non-singular general-relativistic gravitational collapse”, in Proceedings of International Conference GR5, 1968, Tbilisi, USSR, p. 174.

\bibitem{Simpson:2018tsi}
A.~Simpson and M.~Visser,
JCAP \textbf{02}, 042 (2019)
[arXiv:1812.07114 [gr-qc]].



\bibitem{Khodadi:2022dyi}
M.~Khodadi and R.~Pourkhodabakhshi,
Phys. Rev. D \textbf{106} no.8, 084047 (2022)
[arXiv:2210.06861 [gr-qc]].


\bibitem{Sajadi:2023ybm}
S.~N.~Sajadi, M.~Khodadi, O.~Luongo and H.~Quevedo,
Phys. Dark Univ. \textbf{45}, 101525 (2024)
[arXiv:2312.16081 [gr-qc]].



\bibitem{Pedrotti:2024znu}
D.~Pedrotti and S.~Vagnozzi,
[arXiv:2404.07589 [gr-qc]].



\bibitem{Carballo-Rubio:2019fnb}
R.~Carballo-Rubio, F.~Di Filippo, S.~Liberati and M.~Visser,
Phys. Rev. D \textbf{101}, 084047 (2020)
[arXiv:1911.11200 [gr-qc]].


\bibitem{Bonanno:2020fgp}
A.~Bonanno, A.~P.~Khosravi and F.~Saueressig,
Phys. Rev. D \textbf{103}, no.12, 124027 (2021)
[arXiv:2010.04226 [gr-qc]].



\bibitem{Poisson:1989zz}
E.~Poisson and W.~Israel,
Phys. Rev. Lett. \textbf{63}, 1663-1666 (1989)
doi:10.1103/PhysRevLett.63.1663



\bibitem{Balbinot:1993rf}
R.~Balbinot and E.~Poisson,
Phys. Rev. Lett. \textbf{70}, 13-16 (1993).


\bibitem{Bonanno:2022rvo}
A.~Bonanno and F.~Saueressig,
[arXiv:2211.09192 [gr-qc]].



\bibitem{Carballo-Rubio:2018pmi}
R.~Carballo-Rubio, F.~Di Filippo, S.~Liberati, C.~Pacilio and M.~Visser,
JHEP \textbf{07}, 023 (2018)
[arXiv:1805.02675 [gr-qc]].

	
\bibitem{DiFilippo:2022qkl}
F.~Di Filippo, R.~Carballo-Rubio, S.~Liberati, C.~Pacilio and M.~Visser,
Universe \textbf{8}, no.4, 204 (2022)
[arXiv:2203.14516 [gr-qc]].		





\bibitem{Carballo-Rubio:2019nel}
R.~Carballo-Rubio, F.~Di Filippo, S.~Liberati and M.~Visser,
Class. Quant. Grav. \textbf{37}, no.14, 14 (2020)
[arXiv:1908.03261 [gr-qc]].	


\bibitem{Choquet-Bruhat:1969ywq}
Y.~Choquet-Bruhat and R.~P.~Geroch,
Commun. Math. Phys. \textbf{14}, 329-335 (1969).

	
\bibitem{Mellor:1989ac}
F.~Mellor and I.~Moss,
Phys. Rev. D \textbf{41}, 403 (1990).


\bibitem{Brady:1992cz}
P.~R.~Brady, D.~Nunez and S.~Sinha,
Phys. Rev. D \textbf{47}, 4239-4243 (1993)
[arXiv:gr-qc/9211026 [gr-qc]].

\bibitem{Mellor:1992}
F Mellor and I Moss, 
Class. Quantum Grav. 9 L43 (1992).

\bibitem{Brady:1992}
P.~R.~Brady and E.~Poisson,
Class. Quantum Grav. 9 121 (1992).

\bibitem{Chambers:1994ap}
C.~M.~Chambers and I.~G.~Moss,
Class. Quant. Grav. \textbf{11}, 1035-1054 (1994)
[arXiv:gr-qc/9404015 [gr-qc]].


\bibitem{Markovic:1994gy}
D.~Markovic and E.~Poisson,
Phys. Rev. Lett. \textbf{74}, 1280-1283 (1995)
[arXiv:gr-qc/9411002 [gr-qc]].

\bibitem{Cai:1995ux}
R.~G.~Cai and R.~K.~Su,
Phys. Rev. D \textbf{52}, 666-671 (1995).

\bibitem{Cai:1995nt}
R.~G.~Cai,
Phys. Rev. D \textbf{53}, 5698-5704 (1996).

\bibitem{Chambers:1997ef}
C.~M.~Chambers,
Annals Israel Phys. Soc. \textbf{13}, 33 (1997)
[arXiv:gr-qc/9709025 [gr-qc]].

\bibitem{Poisson:1997my}
E.~Poisson,
Annals Israel Phys. Soc. \textbf{13}, 85 (1997)
[arXiv:gr-qc/9709022 [gr-qc]].

\bibitem{Brady:1998au}
P.~R.~Brady, I.~G.~Moss and R.~C.~Myers,
Phys. Rev. Lett. \textbf{80}, 3432-3435 (1998)
[arXiv:gr-qc/9801032 [gr-qc]].

	
\bibitem{Cai:1998yp}
R.~G.~Cai,
Phys. Rev. D \textbf{59}, 104004 (1999)
[arXiv:gr-qc/9810071 [gr-qc]].

\bibitem{Hollands:2019whz}
S.~Hollands, R.~M.~Wald and J.~Zahn,
Class. Quant. Grav. \textbf{37}, no.11, 115009 (2020)
[arXiv:1912.06047 [gr-qc]].


\bibitem{Destounis:2020yav}
K.~Destounis, R.~D.~B.~Fontana and F.~C.~Mena,
Phys. Rev. D \textbf{102}, no.10, 104037 (2020)
[arXiv:2006.01152 [gr-qc]].


\bibitem{Carballo-Rubio:2022kad}
R.~Carballo-Rubio, F.~Di Filippo, S.~Liberati, C.~Pacilio and M.~Visser,
JHEP \textbf{09}, 118 (2022)
[arXiv:2205.13556 [gr-qc]].


\bibitem{Franzin:2022wai}
E.~Franzin, S.~Liberati, J.~Mazza and V.~Vellucci,
Phys. Rev. D \textbf{106}, no.10, 104060 (2022)
[arXiv:2207.08864 [gr-qc]].



\bibitem{Bonanno:2022jjp}
A.~Bonanno, A.~P.~Khosravi and F.~Saueressig,
Phys. Rev. D \textbf{107}, no.2, 024005 (2023)
[arXiv:2209.10612 [gr-qc]].


\bibitem{Carballo-Rubio:2022twq}
R.~Carballo-Rubio, F.~Di Filippo, S.~Liberati, C.~Pacilio and M.~Visser,
[arXiv:2212.07458 [gr-qc]].


\bibitem{Price:1972pw}
R.~H.~Price,
Phys. Rev. D \textbf{5}, 2439-2454 (1972).

\bibitem{Burko:1997tb}
L.~M.~Burko and A.~Ori,
Phys. Rev. D \textbf{56}, 7820-7832 (1997)
[arXiv:gr-qc/9703067 [gr-qc]].


\bibitem{Anderson:1993ni}
W.~G.~Anderson, P.~R.~Brady, W.~Israel and S.~M.~Morsink,
Phys. Rev. Lett. \textbf{70}, 1041-1044 (1993)
[arXiv:gr-qc/9210013 [gr-qc]].


\bibitem{Balbinot:1994ee}
R.~Balbinot and P.~R.~Brady,
Class. Quant. Grav. \textbf{11}, 1763-1773 (1994).

\bibitem{Fulling:1978ht}
S.~A.~Fulling, M.~Sweeny and R.~M.~Wald,
Commun. Math. Phys. \textbf{63}, 257-264 (1978)

\bibitem{Unruh:1976db}
W.~G.~Unruh,
Phys. Rev. D \textbf{14}, 870 (1976).

\bibitem{Hartle:1976tp}
J.~B.~Hartle and S.~W.~Hawking,
Phys. Rev. D \textbf{13}, 2188-2203 (1976).

\bibitem{Javad-Ellis}
 J.~T.~Firouzjaee and G.~F.~R.~Ellis, 
  Phys. Rev. D \textbf{91}, no.10, 103002 (2015)
   [arXiv:1503.05020 [gr-qc]].

\bibitem{Birrell}
N. D. Birrell and P. C. W. Davies, \textit{``Quantum Fields in Curved
Space''}, Cambridge University Press, Cambridge, England,
1984.


\bibitem{Davies:1976ei}
P.~C.~W.~Davies, S.~A.~Fulling and W.~G.~Unruh,
Phys. Rev. D \textbf{13}, 2720-2723 (1976).

\bibitem{Candelas:1980zt}
P.~Candelas,
Phys. Rev. D \textbf{21}, 2185-2202 (1980).

\bibitem{Fawcett:1983dk}
M.~S.~Fawcett,
Commun. Math. Phys. \textbf{89}, 103 (1983).

\bibitem{Balbinot:1999vg}
R.~Balbinot, A.~Fabbri and I.~L.~Shapiro,
Nucl. Phys. B \textbf{559}, 301-319 (1999)
[arXiv:hep-th/9904162 [hep-th]].

\bibitem{McMaken:2023uue}
T.~McMaken,
Phys. Rev. D \textbf{107}, no.12, 125023 (2023)
[arXiv:2303.03562 [gr-qc]].	

\end{thebibliography}
\end{document}